# MuFFIN: Multifaceted Pronunciation Feedback Model with Interactive Hierarchical Neural Modeling

Bi-Cheng Yan, *Student Member*, *IEEE*, Ming-Kang Tsai, and Berlin Chen, *Member*, *IEEE*

***Abstract*—Computer-assisted pronunciation training (CAPT) manages to facilitate second-language (L2) learners to practice pronunciation skills by offering timely and instructive feedback. To examine pronunciation proficiency from multiple facets, existing methods for CAPT broadly fall into two categories: mispronunciation detection and diagnosis (MDD) as well as automatic pronunciation assessment (APA). The former aims to pinpoint phonetic pronunciation errors and provide diagnostic feedback, while the latter seeks instead to quantify pronunciation proficiency pertaining to various aspects. Despite the natural complementarity between MDD and APA, researchers and practitioners, however, often treat them as independent tasks with disparate modeling paradigms. In light of this, we in this paper first introduce MuFFIN, a Multi-Faceted pronunciation Feedback model with an Interactive hierarchical Neural architecture, to jointly address the tasks of MDD and APA. To better capture the nuanced distinctions between phonemes in the feature space, a novel phoneme-contrastive ordinal regularization mechanism is then put forward to optimize the proposed model to generate more phoneme-discriminative features while factoring in the ordinality of the aspect scores. In addition, to address the intricate data imbalance problem in MDD, we design a simple yet effective training objective, which is specifically tailored to perturb the outputs of a phoneme classifier with the phoneme-specific variations, so as to better render the distribution of predicted phonemes meanwhile considering their mispronunciation characteristics. A series of experiments conducted on the Speechocean762 benchmark dataset demonstrates the efficacy of our method in relation to several cutting-edge baselines, showing state-of-the-art performance on both the APA and MDD tasks.**

## I. INTRODUCTION

FUELED by the amplified demand for foreign language acquisition, research on computer-assisted pronunciation training (CAPT) has aroused significant attention amidst the tide of globalization, figuring prominently in the field of computer-assisted language learning (CALL) [1][2]. To bridge the gap between insufficient supplies and pressing needs from language teachers and learners, CAPT systems have emerged as appealing learning tools ubiquitously, shifting the conventional pedagogical paradigm from teacher-led to self-directed learning. Beyond their critical roles in education and language learning, CAPT systems also serve as a handy reference for professionals (e.g., interviewers and examiners) in high-stakes assessments, with the goals of reducing the workload [3][4], alleviating the burdens of recruiting new human experts, and achieving consistent and objective assessment results [5][6][7].

A de-facto archetype system for CAPT is normally instantiated in a read-aloud scenario, where an L2 learner is provided with a reference text and instructed to pronounce it correctly. By taking the learner's speech paired with the reference text as input, CAPT systems are anticipated to assess the learner's oral competence from multiple facets, providing detailed and potentially diagnostic performance feedback with a near-instant turnaround. To this end, mispronunciation detection and diagnosis (MDD) and automatic pronunciation assessment (APA) are two active strands of research in developing pronunciation feedback modules for CAPT. The former seeks to pinpoint phonetic pronunciation errors and provides L2 learners with the corresponding diagnostic feedback [8][9]. The latter, in contrast, concentrates more on assessing the learner's pronunciation quality through multi-faceted pronunciation scores, reflecting his/her proficiency pertaining to specific aspects or some extent of spoken language usage [10][11]. One time-tested approach for MDD is goodness of pronunciation (GOP) and its derivatives [12][13], which calculate the ratio between the likelihoods of the canonical and most likely pronounced phonemes. Phoneme-level erroneous pronunciations are subsequently detected if the likelihood ratios of certain phoneme segments fail below predetermined thresholds. On a separate front, the models of iconic APA methods are typically trained to mimic human ratings based on surface features (viz. a set of hand-crafted features). These models either employ a classifier to predict a holistic score representing learners' oral proficiency [10] or use regressors to estimate continuous analytic scores for specific pronunciation aspects, such as phoneme-level accuracy [14], word-level lexical stress [15], and utterance-level pronunciation quality [16][17].

In spite of the complementary nature of MDD and APA, most existing efforts treat them as independent tasks, thereby developing two disparate feedback modules for use in CAPT. However, some prior studies reveal that an L2 English learner tends to have lower utterance-level assessment scores of intelligibility and fluency [18] whenever his or her utterances frequently contain phoneme-level pronunciation errors [19][20]. In the view of this, we in this paper first propose a novel CAPT modeling paradigm, dubbed MuFFIN, which is a Multi-Faceted pronunciation Feedback model with an Interactive hierarchical Neural structure. MuFFIN unifies the individual feedback modules of MDD and APA into a streamlined, hierarchical neural architecture through a multi-task learning scheme. Building on a language hierarchy aware neural architecture with the tailor-made convolution-augmented Branchformer blocks, MuFFIN can effectively capture interactions across the linguistic granularities (i.e., phoneme, word, and utterance) and preserve fine-grained articulatory cues at different linguistic units. Next, to render the subtle differences between phonemes in the feature space, we introduce a novel phoneme-contrastive ordinal regularizer to facilitate the proposed model in generating more phoneme-discriminative features. This training regime leverages contrastive learning to better align the phoneme representations of a scoring model with the textual embeddings of their corresponding canonical phonemes, while also accounting for the ordinal relationships among the regression targets (i.e., phoneme-level accuracy scores). Furthermore, a simple yet effective training objective, phoneme-specific variation, is explored to ease the data imbalance problem incurred by MDD [21]. Data imbalance is a long-standing problem in MDD, where phoneme distributions are often skewed between correct and incorrect pronunciation instances. As illustrated in Fig. 1, we demonstrate the distributions of vowels with correct and incorrect pronunciations in the training set of the Speechocean762 dataset (see Section IV that follows), which are further categorized into many-shot, medium-shot, and few-shot regions based on their occurrence counts. It is evident that the occurrence of correct pronunciations significantly exceeds that of mispronounced ones. Compounding this issue, correct and incorrect pronunciations exhibit unique long-tailed trends, respectively. For instance, the vowels /EH/ and /EY/ are categorized as medium-shot in correct pronunciations but as many-shot in mispronunciations. Similarly, the vowel /UW/ is found in the many-shot region of correct pronunciations but shifts to the few-shot region of mispronunciations.

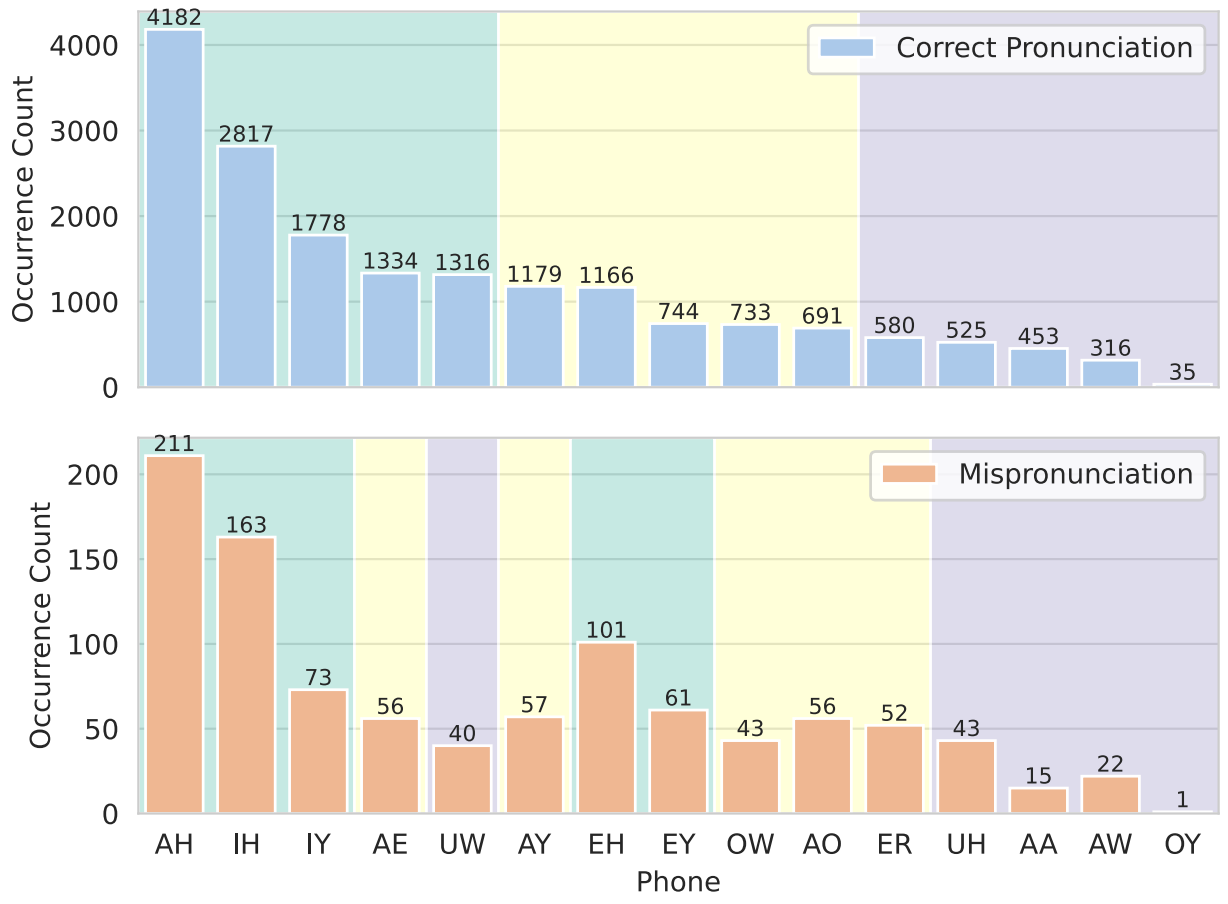

**Fig. 1.** Data imbalance problem of MDD exists in Speechocean762 dataset ( Many-shot,  Medium-shot, and  Few-shot), where the frequencies of correct pronunciations are notably higher than those of mispronunciations. Moreover, the correct and incorrect pronunciations exhibit two distinct long-tailed distributions.

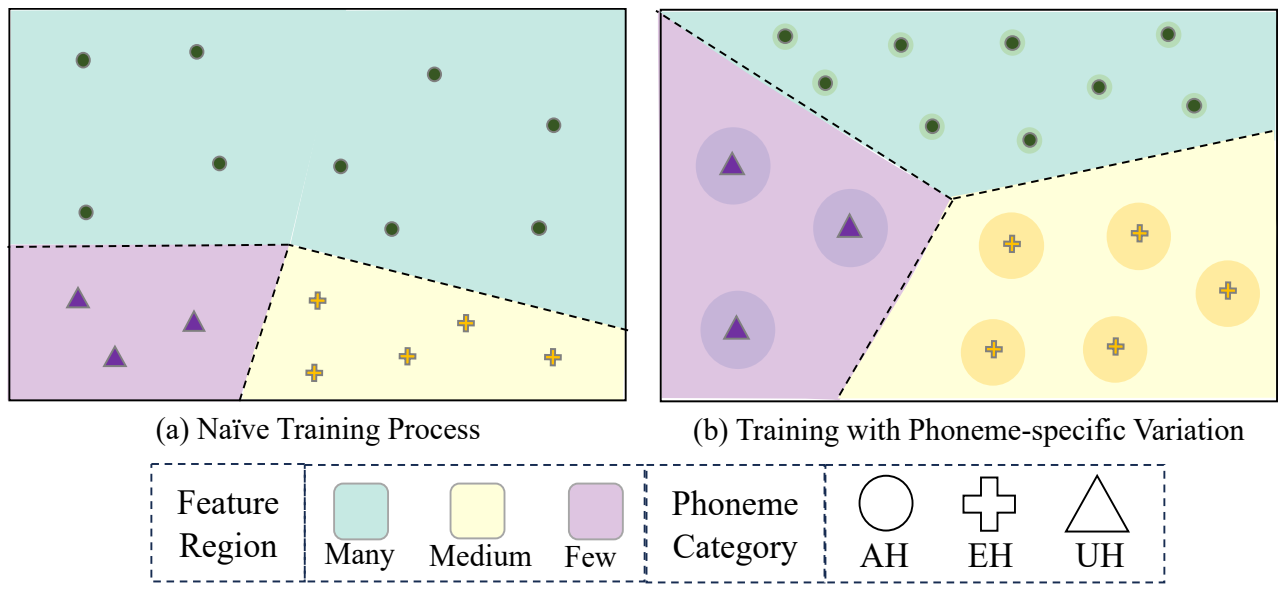

**Fig. 2.** The motivation of the proposed phoneme-specific variation. In the feature space, each point represents a data instance predicted by a phoneme classifier, with different colors indicating distinct categories. (a) The naïve training process tends to bias toward majority phoneme categories, leading to the compression of minority phoneme categories into a narrow area. (b) By applying the proposed phoneme-specific variation training strategy, the feature spaces of minority phoneme categories expand, achieving a more balanced feature distribution while incorporating pronunciation difficulty to modulate feature areas.

Typically, a naïve training process of a phoneme-level pronunciation classifier for MDD is susceptible to an undesirable bias toward correct pronunciations due to their higher occurrence frequency, which therefore dominate the entire training process [22]. As a remedy, our proposed strategy for modeling phoneme-specific variations is built around the hypothesis that logits of phoneme categories with higher occurrence counts (viz. majority phoneme categories) may occupy a larger portion of the feature space, whereas those of with lower occurrence counts (viz. minority phoneme categories) are compressed into a narrower region [23], as depicted in Fig. 2(a). The proposed training strategy augments the logits of phoneme predictions with randomly sampled Gaussian noise, where the radius is determined by the proposed phoneme-specific variation. To address the intricate data imbalance problem of MDD, the modeling of phoneme-specific variations comprises two complementary factors: a quantity factor and a pronunciation difficulty factor. The former assigns smaller variances to majority phoneme categories and larger variances to minority phoneme categories. In contrast, the latter modulates feature areas based on the mispronunciation rates of phonemes. By doing so, as shown in Fig. 2(b), the synergy of the two factors not only balances the feature distributions of disparate phonemes but also adjusts the regions corresponding to pronunciation difficulties. In summary, this paper presents a continuation of our previous work described in [24] and [25] with a significant extension of novel technical contents, experiments, and analysis, whose main contributions are at least four-fold:

- We present MuFFIN, a multi-faceted pronunciation feedback model that jointly addresses the tasks of MDD and APA through an interactive hierarchical neural framework.

This model signifies a paradigm shift from separate modeling of APA and MDD to a unified assessment approach, opening up a new avenue in CAPT.

- A contrastive phonetic ordinal regularizer is proposed to align the speech-derived phoneme representations with the corresponding phoneme-level textual embeddings, while organically engaging the ordinality of pronunciation accuracy scores. A series of graphical examinations are conducted through the lens of the ordinality and phoneme properties.
- To the best of our knowledge, this is the first attempt to address data imbalance issues in MDD by incorporating phoneme-specific variations into the training process. Our method highlights that the data imbalance problem in MDD stems from two intertwined and equally crucial factors, viz. the quantity and the pronunciation difficulty of the training data. Our empirical findings reveal that addressing the data-imbalance problem of MDD by solely considering the data quantity factor primarily enhances the recall metric but sacrifices the precision metric. This analysis has led us to propose a training strategy that incorporates a pronunciation difficulty factor, achieving a better balance between recall and precision metrics compared to strategies that consider either the quantity factor or the pronunciation difficulty factor individually.
- Extensive sets of experiments carried out on the Speechocean762 benchmark dataset [26] confirm the effectiveness of our proposed methods, which improves the performance of state-of-the-art ones on both the APA and MDD tasks.

## II. Related Work

Computer-assisted pronunciation training (CAPT) is a subfield of computer-assisted language learning (CALL), whose research and development can trace back to pioneering efforts in the 1960s [27] and have gained significant attention recently due to the unprecedented advancements in speech and language technologies [28][29][30]. According to the diagnostic feedback of CAPT, research endeavors typically fall into phoneme-level mispronunciation detection and diagnosis (MDD) as well as automatic pronunciation assessment (APA), both mostly developed under read-aloud learning scenarios.

### *A. Mispronunciation Detection and Diagnosis*

Mispronunciation detection and diagnosis (MDD) manages to detect erroneous pronunciation at phoneme segments, and in turn provide L2 learners with the corresponding diagnostic feedback [31][32]. Common approaches to MDD can be grouped into three categories: pronunciation scoring-based, dictation-based, and prompt-based methods. Pronunciation scoring-based methods typically exploit various types of confidence measurements to evaluate pronunciation quality via a well-trained ASR system (e.g., hybrid DNN-HMM ASR system). Frequently-used measurements include, but are not limited to, phoneme durations [33][34], likelihood ratios [13], phoneme posterior probabilities [35], and their combinations [36]. Given an input utterance and its corresponding canonical phoneme sequence (viz. phoneme-level text prompt), pronunciation scoring-based methods first gauge the pronunciation scores for each phoneme in the canonical phoneme sequence. Mispronounced phoneme segments are then detected when their scores fall below predetermined thresholds, signifying a deviation from the expected pronunciation. However, pronunciation scoring-based methods are untenable to provide diagnostic feedback for the detected mispronounced phoneme segments. As a remedy, dictation-based methods strive to formulate MDD as a phoneme recognition task by employing a phoneme recognizer to dictate the most likely phoneme sequence uttered by an L2 learner. The erroneous pronunciation portions can be easily identified by comparing the dictation result with the corresponding canonical phoneme sequence. For instance, Leung et al. ventured into employing a CTC-based phoneme recognizer for L2 English learners, showing comparative performance with pronunciation scoring-based methods in the mispronunciation detection subtask, where the performance gains mainly contributed from the accurate diagnosis of mispronunciations in unvoiced phoneme segments [37]. Yan et al. exploited the hybrid CTC-Attention ASR model as the dictation model and sought to capture deviant (non-categorical) phoneme productions uttered by accented L2 learners with anti-phone modeling [38]. Both of the above-mentioned methods rely on precise alignments to identify mispronounced segments; however, in practical applications, alignment errors might arise when comparing the canonical phoneme sequence to accented or disfluent speech produced by L2 learners. In response, prompt-based methods leverage an attention mechanism to derive the soft alignment between the canonical phoneme sequence and the learner's input speech in an end-to-end manner, offering a promising approach to reduce alignment errors. As one of the first attempts, PeppaNet aligns canonical phonemes with the learner's speech via a Transformer decoder, where any discrepancies are captured in the matching degree vectors through end-to-end neural modeling [39]. Among other things, MDDGCN introduces a graph-based prompt encoder for canonical phonemes, aiming to improve diagnosis accuracy by regularizing the relationships between canonical and actually pronounced phonemes through a pre-defined phonetic graph [40].

### *B. Automatic Pronunciation Assessment*

Automatic Pronunciation Assessment (APA) quantifies an L2 learner's pronunciation proficiency in a target language by providing either analytic scores (viz. continuous numerical values) for specific pronunciation aspects [41][42] or a holistic assessment (viz. discrete categorical values) to reflect overall spoken competence [10]. Early efforts in APA predominantly focused on single-aspect assessment, typically by constructing individual scoring modules to predict proficiency scores at specific linguistic levels with various sets of hand-crafted features. These hand-crafted features, extracted from the learner's input speech or its corresponding ASR-generated

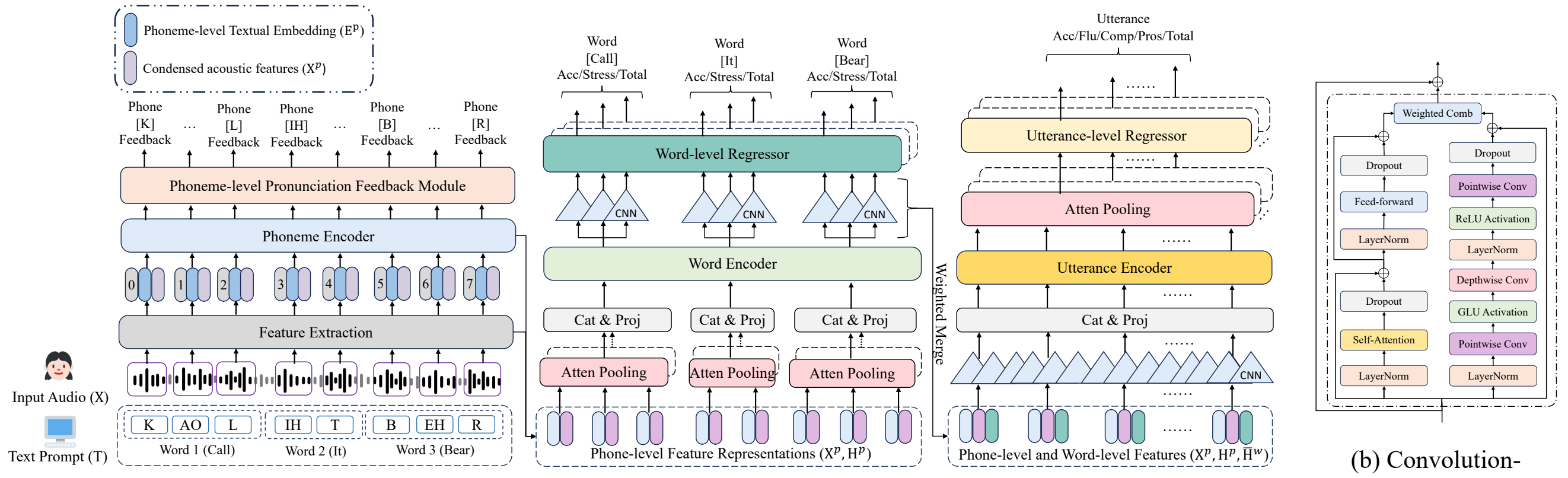

(a) Multi-faceted Pronunciation Feedback Model with an Interactive Hierarchical Neural Architecture (MuFFIN)   (b) Convolution-augmented Branchformer transcript, may include acoustic features, confidence scores of recognized linguistic units, time-alignment information, and statistical measures [43][44]. To scrutinize learners' pronunciation comprehensively, recent advances in APA have advocated multi-aspect and multi-granular pronunciation assessment, leveraging unified scoring models that evaluate pronunciation proficiency across multiple linguistic levels (viz. phoneme, word, and utterance) with diverse aspects (e.g., accuracy, fluency, and completeness). Drawing on this research tend, Gong et al. proposed a parallel pronunciation modeling architecture dubbed GOPT, which took GOP features as input and adopted a Transformer encoder as the backbone model to jointly model multiple pronunciation aspects across various linguistic granularities [45]. Following this school of thought, 3M extended GOPT by augmenting the model's input embeddings with prosodic features and self-supervised learning (SSL)-based features, aiming to achieve multi-view, multi-granularity, and multi-aspect pronunciation modeling [46]. Despite their decent performance, the hierarchical structure of an utterance is largely set aside. To capture the language hierarchy of an utterance, Do et al. proposed a hierarchical APA model and explored a novel multi-trait attention layer to strengthen the connection between scoring aspects [47]. Chao et al. introduced sub-phoneme modeling and employed a depth-wise separable convolution layer to construct a hierarchical APA model, facilitating better modeling of local context cues at the sub-word level [48]. Apart from the above, Gradformer leveraged a granularity-decoupled Transformer network that first separates the granularity of an utterance into lower-level (phoneme- and word-level) ones and higher-level (utterance-level) one. A Conformer encoder in turn jointly models pronunciation aspects at the lower levels, while a Transformer decoder processes a set of trainable aspect vectors and interacts with the encoder outputs for utterance-level pronunciation assessment [42].

**Fig. 3.** The proposed multi-faceted pronunciation feedback model with an interactive hierarchical neural architecture (MuFFIN). (a) The overall model architecture processes input audio and text prompts, hierarchically representing the learner's pronunciation to generate assessment scores across various aspects. (b) The proposed convolution-augmented Branchformer block functions as the backbone of MuFFIN, operating in encoders at different linguistic levels. The pronunciation aspects of accuracy, fluency, completeness, and prosody are denoted as Acc, Flu, Comp, and Pros, respectively.

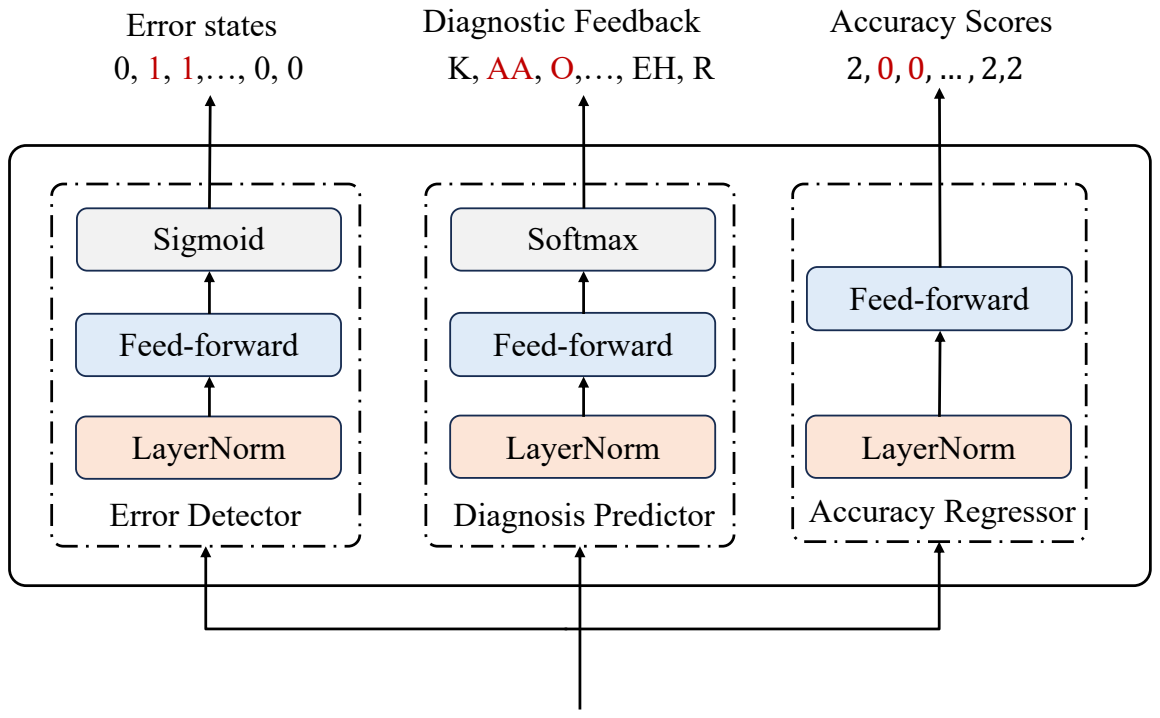

**Fig. 4.** The proposed phoneme-level pronunciation feedback module, which takes the output of phoneme encoder and simultaneously generate phoneme-level error states, diagnostic phonemes, and accuracy scores.

## III. MULTI-FACETED PRONUNCIATION FEEDBACK MODEL WITH AN INTERACTIVE HIERARCHICAL NEURAL ARCHITECTURE

The overall architecture of the proposed MuFFIN is schematically depicted in Fig. 3(a), which contains three main components: phoneme-level modeling, word-level modeling, and utterance-level modeling. The encoder at each different linguistic level adopts a novel convolution-augmented Branchformer block [25], as shown in Fig. 3(b), which consists of two branches with one branch designed to capture supra-segmental pronunciation cues with multi-head attention (MHA) layers while the other tailored to capture fine-grained pronunciation cues with a series of convolution layers. Furthermore, as illustrated in Fig. 4, a novel phoneme-level pronunciation feedback module is devised to assess phoneme-level accuracy and perform mispronunciation detection and diagnosis.

### *A. Problem Formulation*

Given an input utterance U, consisting of a time sequence of audio signals X produced by an L2 learner, and a reference text prompt T with $M$ words, which is converted into $N$ canonical phonemes based on a pronunciation dictionary[1], the proposed multi-faceted pronunciation feedback model aims to estimate proficiency scores at various linguistic granularities, while pinpointing phoneme-level pronunciation errors for the canonical phoneme sequence. Formally, let $\mathrm{G} = \{p, w, u\}$ be a set of linguistic granularities, where $p, w, u$ stands for the phoneme-, word-, and utterance-level, respectively. For each granularity $g \in G$, our model aims to predict a set of aspect score sequences $\mathrm{A}^g = \left\{\mathbf{a}_1^g, \mathbf{a}_2^g, \dots, \mathbf{a}_{N_g}^g\right\}$, where $N_g$ is the number of pronunciation aspects at granularity $g$. In the meantime, for the canonical phoneme sequence $\mathbf{q} = (q_1, q_2, \dots, q_N)$, the proposed model seeks to detect an error state sequence $\mathbf{e} = (e_1, e_2, \dots, e_N)$ and generate a phonetic diagnosis sequence $\mathbf{y} = (y_1, y_2, \dots, y_N)$. Both $e_n$ and $y_n$ are phoneme-level pronunciation feedback for $q_n$, where $e_n = 1$ denotes a mispronounced phoneme segment and $e_n = 0$ for a correct one, while $y_n$ specifies the phoneme produced by the learner.

### *B. Interactive Hierarchical Neural Modeling*

**Phoneme-level Modeling.** For an input utterance, we first extract various pronunciation features to portray the pronunciation quality of the L2 learner at phoneme-level, which are then concatenated and projected to obtain a sequence of condensed acoustic features $\mathrm{X}^p$. The feature extraction process is formulated as:

$$\mathrm{X}^p = \mathrm{Linear}_p([\mathrm{E}^{\mathrm{GOP}}; \mathrm{E}^{\mathrm{Dur}}; \mathrm{E}^{\mathrm{Eng}}; \mathrm{E}^{\mathrm{SSL}}]), \tag{1}$$

where $\mathrm{Linear}_p(\cdot)$ is a single feedforward layer, $\mathrm{E}^{\mathrm{GOP}}$ is goodness of pronunciation (GOP)-based features including log phoneme posterior (LPP) and log posterior ratio (LPR) [12][14]. $\mathrm{E}^{\mathrm{Dur}}$ and $\mathrm{E}^{\mathrm{Eng}}$ are prosodic features related to duration and energy statistics [49][50], while $\mathrm{E}^{\mathrm{SSL}}$ are self-supervised learning (SSL) based features [46]. We then add phoneme-level textual embeddings $\mathrm{E}^p$ to $\mathrm{X}^p$, followed by a phoneme encoder to obtain aspect representations $\mathrm{H}^p = (\mathbf{h}_1^p, \mathbf{h}_2^p, \dots, \mathbf{h}_N^p)$:

$$\mathrm{H}_0^p = \mathrm{X}^p + \mathrm{E}^p, \tag{2}$$

$$\mathrm{H}^p = \mathrm{PhnEnc}(\mathrm{H}_0^p). \tag{3}$$

Here, $\mathrm{E}^p$ is generated by passing $\mathbf{q}$ into a phoneme-level prompt encoder which comprises a phoneme and position embedding layer. $\mathrm{PhnEnc}(\cdot)$ is composed of a stack of 3 convolution-augmented Branchformer blocks.

Afterward, the pronunciation feedback module builds on $\mathrm{H}^p$ to estimate the multi-faceted pronunciation feedback, comprising three components: an error detector, a diagnosis predictor, and an accuracy score regressor. The error detector is a binary labeling model which predicts the error state $\hat{e}_n$, indicating whether the $n$-th phoneme of $\mathbf{q}$ is identified as a mispronunciation:

$$\mathrm{P}_{det}(\hat{e}_n|\mathbf{q}, \mathrm{X}) = \mathrm{Sigmoid}\big(\mathrm{Linear}_{det}(\mathbf{h}_n^p)\big), \tag{4}$$

where $\mathrm{Linear}_{det}(\cdot)$ is a linear layer followed by layer normalization. The diagnostic predictor performs a sequential multi-class labeling process to derive the probability distribution of diagnostic feedback for the $n$-th canonical phoneme as:

$$\mathrm{P}_{diag}(\hat{y}_n|\mathbf{q}, X) = \mathrm{Softmax}\Big(\mathrm{Linear}_{diag}(\mathbf{h}_N^p)\Big), \tag{5}$$

where $\mathrm{Linear}_{diag}(\cdot)$ used to convert hidden dimensions into the size of pronunciation dictionary. Finally, the phoneme-level accuracy score is estimated by an accuracy score regressor.

**Word-level Modeling.** For the word-level assessments, a word-level attention pooling is introduced to produce a word representation vector from its constituent phonemes, instantiated with a 1-D depth-wise convolution layer followed by an MHA layer and an average operation. The word-level input representations $\mathrm{X}^w$ are computed by individually passing $\mathrm{X}^p$ and $\mathrm{H}^p$ into the word-level attention pooling and subsequently packing them together with a linear projection:

$$\widehat{\mathrm{X}}^w, \widehat{\mathrm{H}}^w = \mathrm{AttPool}_{w_1}(\mathrm{X}^p), \mathrm{AttPool}_{w_2}(\mathrm{H}^p), \tag{6}$$

$$\mathrm{X}^w = \mathrm{Linear}_w\Big(\big[\widehat{\mathrm{X}}^w; \widehat{\mathrm{H}}^w\big]\Big). \tag{7}$$

Next, the word-level textual embeddings $\mathrm{E}^w$ are added to $\mathrm{X}^w$, and a word encoder is employed to generate word-level contextualized representations $\mathrm{H}^w$:

$$\mathrm{H}_0^w = \mathrm{X}^w + \mathrm{E}^w, \tag{8}$$

$$\mathrm{H}^w = \mathrm{WordEnc}(\mathrm{H}_0^w), \tag{9}$$

where $\mathrm{E}^w$ are obtained by mapping the text prompt T into the corresponding embedding sequence via a word and position embedding layer, and $\mathrm{WordEnc}(\cdot)$ consists of 2 convolution-augmented Branchformer blocks. Finally, three distinct 1-D depth-wise convolution layers are performed on $\mathrm{H}^w$ to generate word-level aspect representations (i.e., $\mathrm{H}^{w_1}, \mathrm{H}^{w_2}$, and $\mathrm{H}^{w_3}$), which are then transformed into the pronunciation score sequences by the corresponding word-level regressors.

**Utterance-level Modeling.** For the utterance-level assessments, we first extract an utterance-level SSL-based feature $\overline{\mathrm{E}}^{\mathrm{SSL}}$ by applying average pooling over the time dimension of frame-level SSL-based features. Next, we merge $\mathrm{H}^{w_1}, \mathrm{H}^{w_2}$, and $\mathrm{H}^{w_3}$ with a weighted combination to obtain word-level representations $\overline{\mathrm{H}}^w$. A sequence of utterance-level input representations $\mathrm{H}_0^u$ is obtained by first applying 1-D depth-wise convolution layers to $\mathrm{X}^p$, $\mathrm{H}^p$, and $\overline{\mathrm{H}}^w$, followed by concatenation and linear projection. Consequently, an utterance encoder is exploited to generate contextualized representations $\mathrm{H}^u$:

$$\overline{\mathrm{H}}^w = \mathrm{Merge}(\mathrm{H}^{w_1}, \mathrm{H}^{w_2}, \mathrm{H}^{w_3}), \tag{10}$$

$$\mathrm{H}_0^u = \mathrm{Linear}_u([\mathrm{DC}_1(\mathrm{X}^p); \mathrm{DC}_2(\mathrm{H}^p); \mathrm{DC}_3(\overline{\mathrm{H}}^w)]), \tag{11}$$

$$\mathrm{H}^u = \mathrm{UttEnc}(\mathrm{H}_0^u), \tag{12}$$

where $\mathrm{Merge}(\cdot)$ is a weighted average operation [51], $\mathrm{UttEnc}(\cdot)$ is a single convolution-augmented Branchformer block, and $\mathrm{DC}_1(\cdot), \mathrm{DC}_2(\cdot), \mathrm{DC}_3(\cdot)$ are distinct 1-D depthwise convolution layers, each with a kernel size of 3. Afterward, five separate attention pooling modules are built on top of $\mathrm{H}^u$ to generate utterance-level aspect representation vectors. These features are then combined with $\overline{\mathrm{E}}^{\mathrm{SSL}}$ via the residual connections and converted into utterance-level aspect scores through the respective regressors.

[1] CMU dictionary: http://www.speech.cs.cmu.edu/cgi-bin/cmudict

**Training Objective.** The training objective of the proposed model is calculated from the losses of APA and MDD:

$$\mathcal{L}_{MuFFIN} = \mathcal{L}_{APA} + \mathcal{L}_{MDD}. \quad (13)$$

The APA loss is a weighted sum of the mean square error (MSE) losses gathered from different granularity levels:

$$\mathcal{L}_{APA} = \sum_{j_p} \frac{\mathcal{L}_{p^{j_p}}}{N_p} + \sum_{j_w} \frac{\mathcal{L}_{w^{j_w}}}{N_w} + \sum_{j_u} \frac{\mathcal{L}_{u^{j_u}}}{N_u}, \quad (14)$$

where $\mathcal{L}_{p^{j_p}}$, $\mathcal{L}_{w^{j_w}}$, and $\mathcal{L}_{u^{j_u}}$ are phoneme-level, word-level, and utterance-level losses for disparate aspects, and $N_p$, $N_w$, $N_u$ mark the numbers of aspects at each granularity. On a separate front, the training objective of MDD comes with the tasks of mispronunciation detection $\mathcal{L}_{det}$ and diagnosis $\mathcal{L}_{diag}$:

$$\mathcal{L}_{MDD} = \mathcal{L}_{det} + \mathcal{L}_{diag}, \quad (15)$$

$$\mathcal{L}_{det} = -\sum_{n=1}^{N} \log \mathrm{P}_{det}(\hat{e}_n = e_n | \mathbf{q}, \mathrm{X}), \quad (16)$$

$$\mathcal{L}_{diag} = -\sum_{n=1}^{N} \log \mathrm{P}_{diag}(\hat{y}_n = y_n | \mathbf{q}, \mathrm{X}), \quad (17)$$

where $\mathcal{L}_{det}$ and $\mathcal{L}_{diag}$ represent the negative log-likelihood used for training the detector and the predictor, respectively.

## IV. Contrastive Phonemic Ordinal Regularizer

To generate more phoneme-discriminative features for the multi-faceted pronunciation assessment model, we proposed contrastive phonemic ordinal regularizer (ConPCO), which consists of three mathematical terms: the contrastive term $\mathcal{L}_{con}$, the phonemic characteristic term $\mathcal{L}_{pc}$, and the ordinal term $\mathcal{L}_o$. $\mathcal{L}_{con}$ aims to simultaneously project the phoneme representations generated from a pronunciation assessment model and the embeddings of phoneme-level text prompt into a joint feature space. $\mathcal{L}_{pc}$ and $\mathcal{L}_o$ adjust the distances between inter- and intra-phoneme categories, where the former enhances inter-phoneme discrepancy, and the latter improves intra-phoneme compactness with ordinal relationship. The proposed ConPCO regularizer is formulated as:

$$\mathcal{L}_{ConPCO} = \mathcal{L}_{con} + \mathcal{L}_{pc} + \mathcal{L}_o. \quad (18)$$

**Contrastive Term.** Let $\mathrm{H}^p = (\mathbf{h}_1^p, \mathbf{h}_2^p, \dots, \mathbf{h}_N^p)$ stand for the phoneme representation sequence of an utterance generated by a phoneme encoder in a pronunciation scoring model, and $\mathrm{E}^p = (\mathbf{e}_1^p, \mathbf{e}_2^p, \dots, \mathbf{e}_N^p)$ denote the textual embedding of canonical phonemes generated by a phoneme-level prompt encoder. Next, a set of paired phoneme representations $\mathcal{M} = \{(\mathbf{z}_i^p, \mathbf{z}_i^t), i = 1, \dots, M\}$ is obtained by first applying separate linear projections to $\mathrm{H}^p$ and $\mathrm{E}^p$ , and then calculating the centroid vectors for each phoneme category. Next, as illustrated in Fig. 5, the $M \times M$ similarities are derived from $\mathcal{M}$ , with the contrastive term $\mathcal{L}_{con}$ aiming to maximize the similarity between paired phoneme representations while minimizing the similarity of unpaired ones [52][53]. The contrastive term $\mathcal{L}_{con}$ includes two losses, with a temperature hyper-parameter $\tau$ that controls the strength of penalties on negative samples:

$$\mathcal{L}_{con} = \mathcal{L}_{p2t} + \mathcal{L}_{t2p}, \quad (19)$$

$$\mathcal{L}_{p2t} = -\frac{1}{M}\sum_{i=1}^{M} \log \frac{\exp(\phi(\mathbf{z}_i^p, \mathbf{z}_i^t)/\tau)}{\sum_{j=1}^{M} \exp\left(\phi\left(\mathbf{z}_i^p, \mathbf{z}_j^t\right)/\tau\right)}, \quad (20)$$

$$\mathcal{L}_{t2p} = -\frac{1}{M}\sum_{i=1}^{M} \log \frac{\exp(\phi(\mathbf{z}_i^t, \mathbf{z}_i^p)/\tau)}{\sum_{j=1}^{M} \exp\left(\phi\left(\mathbf{z}_i^t, \mathbf{z}_j^p\right)/\tau\right)}, \quad (21)$$

where $\phi(\mathbf{z}_i^p, \mathbf{z}_j^t)$ is dot product between $\ell_2$ normalized vectors $\mathbf{z}_i^p$ and $\mathbf{z}_j^t$ (cosine similarity). During training, $\mathcal{M}$ is constructed from each batch, where we empirically sample the data instances with the highest proficiency score to compute centroid vectors.

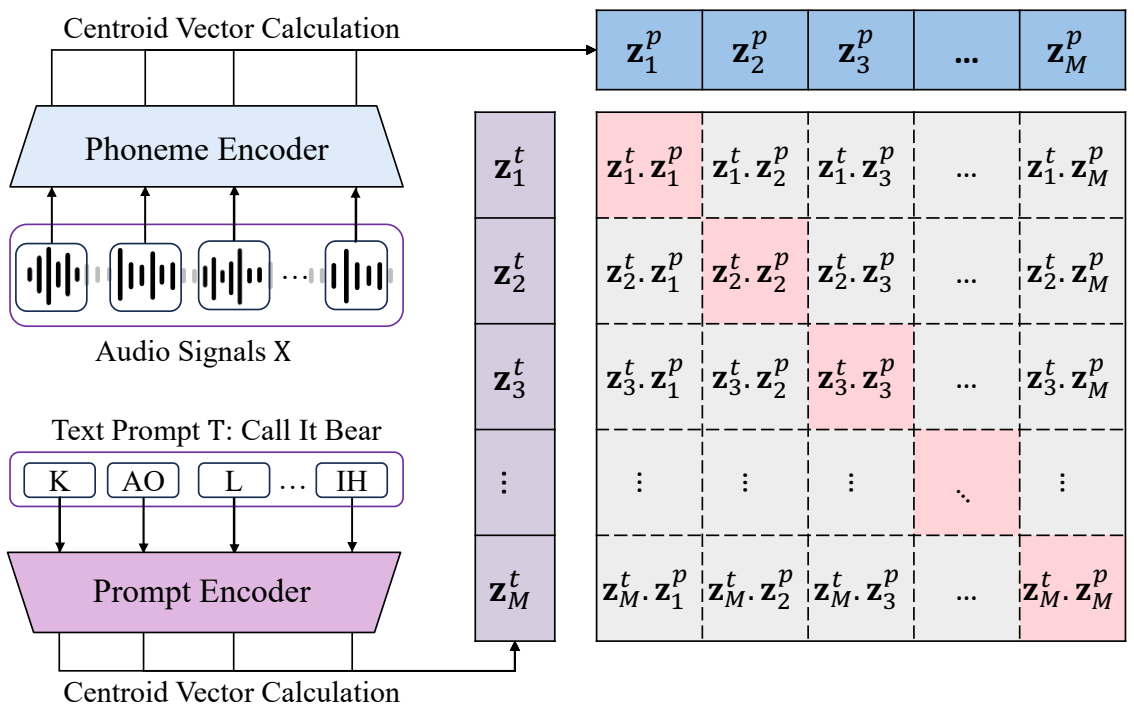

**Fig. 5.** The visualization of the calculation process for the contrastive term $\mathcal{L}_{con}$.

**Phonemic Characteristic Term.** The phonemic characteristic term $\mathcal{L}_{pc}$ preserve the phonemic proximity information by minimize the negative distances between centroid vectors $\mathbf{z}_i^p$:

$$\mathcal{L}_{pc} = -\frac{1}{M(M-1)}\sum_{i=1}^{M}\sum_{i \neq j} \left\| \mathbf{z}_i^p - \mathbf{z}_j^p \right\|_2, \quad (22)$$

where $\mathcal{L}_{pc}$ is equivalent to maximizing the distances between phoneme categories during the optimization process.

**Ordinal Term.** To reflect ordinal relationships of regression targets in the feature space, the ordinal term $\mathcal{L}_o$ is defined to minimize the distance between the feature representations $\mathbf{h}_i^p$ and their corresponding phoneme centroid vectors $\mathbf{z}_i^p$ with relative differences of proficiency score:

$$\mathcal{L}_o = \frac{1}{N}\sum_{i=1}^{N} w_i \|\mathbf{h}_i^p - \mathbf{z}_i^p\|_2, \quad (23)$$

where $w_i = |C - y_i^p|$ is a compactness weight for each $\mathbf{h}_i^p$, reflecting the ordinal behaviors within the label space, with $y_i^p$ denoting the corresponding phoneme-level accuracy score. The tunable constant $C$ is set to be 3, representing the highest accuracy score plus a small margin.

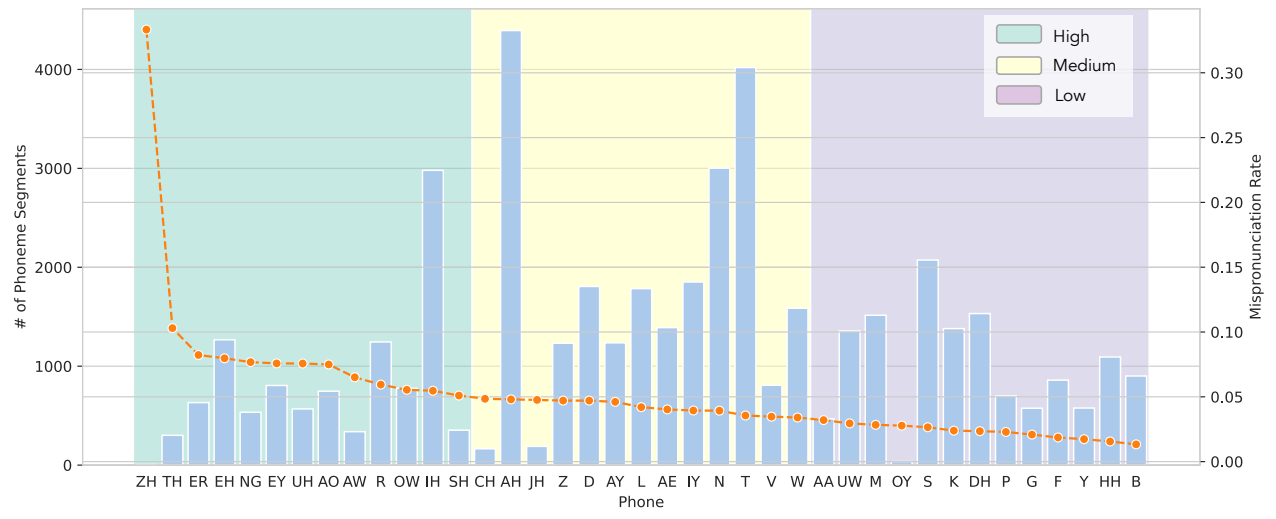

**Fig. 6.** Phoneme statistics of Speechocean762, including occurrence count and corresponding mispronunciation rate.

## IV. Phoneme-specific Variation

To balance the distribution of predicted phonemes while accounting for pronunciation difficulties, the logits of phoneme predictions generated by a phoneme predictor are perturbed with randomly sampled Gaussian noise, where the radius is determined by the phoneme-dependent variance. To this end, the proposed training scheme, phoneme-specific variation (PhnVar), consists of two factors, i.e., a data quantity factor and a pronunciation difficulty factor. The data quantity factor assigns smaller variance to majority phoneme categories and larger variance to minority ones, while the pronunciation difficulty factor modulates feature areas based on mispronunciation rates. Formally, we revisit Eq. (5) and express the probability of the $n$-th canonical phoneme being predicted as a diagnostic phoneme $k$, derived from the softmax function:

$$p_k^n = \frac{\exp(g_k^n)}{\sum_{i=1}^{M} \exp(g_i^n)}. \tag{24}$$

Here, $g_k^n$ is the logit of the $k$-th phoneme in logit vector $\mathbf{g}^n = (g_1^n, g_2^n, \ldots, g_M^n)$, generated by $\mathrm{Linear}_{\mathrm{diag}}(\mathbf{h}_N^p)$, where $M$ is the number of phoneme categories. We then augment logits with phoneme-specific variance, defined as the weighted p of the data quantity factor $\mathrm{QF}_k$ and the pronunciation difficulty factor $\mathrm{DF}_k$ for phoneme $k$, with coefficients $\alpha$ and $\beta$:

$$\hat{g}_k^n = g_k^n + \delta(\sigma) \times \exp\left(\frac{\alpha \times \log(\mathrm{QF}_k) + \beta \times \log(\mathrm{DF}_k)}{\alpha + \beta}\right), \tag{25}$$

where $\delta(\sigma)$ stands for a Gaussian distribution with a zero mean and the standard deviation $\sigma$. Both $\alpha$ and $\beta$ are set to 1 in our experiments. The data quantity factor is defined as normalized inverse phoneme frequency operated in the logarithmic scale:

$$\mathrm{QF}_k = \frac{c_k}{\max_i c_i}; \quad c_k = \log \frac{\sum_{i=1}^{M} q_i}{q_k}, \tag{26}$$

where $q_k$ is the number of instances in phoneme category $k$. The pronounce difficulty factor is expressed as normalized mispronunciation rate:

$$\mathrm{DF}_k = \frac{d_k}{max_i\, d_i}; \quad d_k = \frac{mp_k}{mp_k + cp_k}, \tag{27}$$

where $mp_k$ and $cp_k$ are the number of mispronounced and correctly pronounced instances for phoneme category $k$, respectively.

TABLE I
STATISTICS OF APA TASK IN THE SPEECHOCEAN762

| Automatic Pronunciation Assessment | | | | |
|---|---|---|---|---|
| Granularity | Aspect | Score Interval | # of Counts | |
| | | | Train | Test |
| Phoneme | Accuracy | [0, 2] | 47,076 | 47,369 |
| Word | Accuracy<br>Stress<br>Total | [0, 10] | 15,849 | 15,967 |
| Utterance | Accuracy<br>Completeness<br>Fluency<br>Prosody<br>Total | [0, 10] | 2,500 | 2,500 |

TABLE II
STATISTICS OF MDD TASK IN THE SPEECHOCEAN762

| Mispronunciation Detection and Diagnosis | | | |
|---|---|---|---|
| Type | Description | Counts | |
| | | Train | Test |
| Correctness | The uttered phoneme aligns with the canonical phoneme | 45,088 | 45,959 |
| Deletion | A canonical phoneme is omitted | 450 | 396 |
| Substitution | A canonical phoneme is mispronounced to others | 914 | 593 |
| Non-categorical Error | The uttered phoneme not exists in the CMU pronunciation dictionary | 488 | 332 |
| Accented Error | A canonical phoneme is pronounced correctly but with a strong accent | 136 | 89 |

## IV. Experimental Setups

### A. Experimental Data and Evaluation Metrics

**Dataset.** A series of experiments were conducted on the Speechocean762 dataset, a publicly available dataset specifically designed for research on computer-assisted language learning [26]. This dataset contains 5,000 English-speaking recordings spoken by 250 Mandarin L2 learners. The training and test sets are of equal size, and each of them has 2,500 utterances. For the APA task, pronunciation proficiency scores were evaluated at multiple linguistic granularities with various pronunciation aspects, as the statistics of the APA task are summarized in Table I. For the MDD task, the phoneme labels follow the definitions in the CMU pronunciation dictionary, which includes a set of 39 canonical phonemes. In Speechocean762, mispronunciation labels were manually assigned to phoneme segments with accuracy scores below 0.5 and were categorized into four types: deletion, substitution, non-categorical error, and accented error. Table II summarizes the phoneme segment statistics for the MDD task.

**Evaluation Metrics.** The primary evaluation metric for APA adopts Pearson correlation coefficient (PCC), which measures the linear correlation between predicted scores and ground-truth

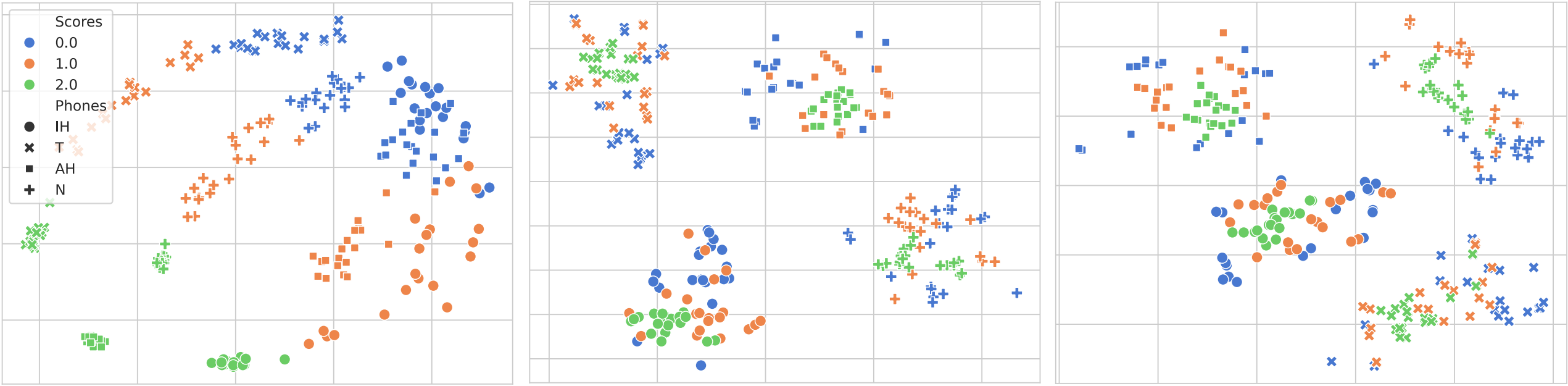

(a) MuFFIN (b) MuFFIN + $\mathcal{L}_{pc}$ (c) MuFFIN $+\mathcal{L}_{pc}+\mathcal{L}_{o}$

**Fig. 7.** Visualization of phoneme representations from MuFFIN trained with the phonemic characteristic and the ordinal terms (i.e., $\mathcal{L}_{pc}$ and $\mathcal{L}_{o}$). For each data point, color indicates the phoneme-level accuracy score, and shape denotes the corresponding phoneme category. We show the phoneme representations $\mathrm{H}^{p}$ of (a) the vanilla MuFFIN model, (b) MuFFIN+$\mathcal{L}_{pc}$, and (c) MuFFIN+$\mathcal{L}_{pc}+\mathcal{L}_{o}$.

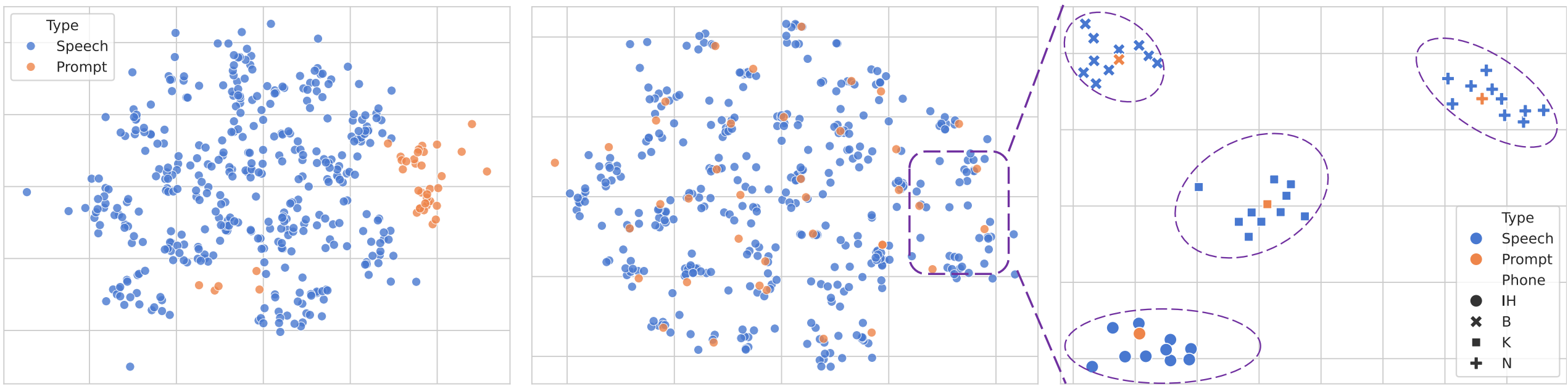

(a) MuFFIN (b) MuFFIN + $\mathcal{L}_{con}$ (c) Localized Zoomed-in Counterpart

**Fig. 8.** Visualization of phoneme representations, with the blue and orange points denoting feature representations $\mathrm{H}^{p}$ and $\mathrm{E}^{p}$, generated from the phoneme encoder of MuFFIN and the phoneme-level prompt encoder, respectively. The feature representations are displayed for (a) the vanilla MuFFIN model, (b) MuFFIN + $\mathcal{L}_{con}$, and (c) a localized zoomed-in counterpart.

scores. In accordance with prior studies, mean square error (MSE) is reported for phoneme-level accuracy. On the other hand, for MDD tasks, the evaluation metrics follow the scoring rubrics in [9]. Specifically, the mispronunciation detection subtask is evaluated using recall (RE), precision (PR), and F1-score (F1), while the mispronunciation diagnosis subtask is assessed with diagnostic error rate (DER), false rejection rate (FRR), false acceptance rate (FAR), and phoneme error rate (PER).

*B. Implementation Details*

**Feature Extraction.** For the pronunciation feature extraction, the GOP features, the energy, and the duration statistics are adopted in line with our previous studies [24][25]. The extraction of SSL-based features follows the processing flow suggested in [46], where features are extracted from the outputs of pretrained acoustic models, including Wav2vec2.0 [54], WavLM [55], and HuBERT [56]. The SSL-based and energy features are extracted at the frame level and then aggregated into phoneme-level representations based on timestamps of phoneme segments derived from forced-aligning the learner's speech to the reference text. The extracted phoneme-level proficiency features amount to 3,164 dimensions, comprising 84 dimensions for GOP features $\mathrm{E}^{\mathrm{GOP}}$, 7 for energy statistic $\mathrm{E}^{\mathrm{Eng}}$, 1 for duration value $\mathrm{E}^{\mathrm{Dur}}$, and 3,072 for SSL-based features $\mathrm{E}^{\mathrm{SSL}}$.

**Training Configuration.** For the training configuration, we followed to the settings reported in [24][25], where each experiment consisted of 5 independent trials, and each trial runs for 100 epochs with different random seeds. In each trial, the model was trained with an Adam optimizer with an initial learning rate of 1e-3 and a batch size of 25. A learning rate scheduler was used to decay the learning rate by a factor of 0.1 after the overall loss did not decrease for 10 consecutive epochs. Furthermore, our models were initialized with a pretrained model following the pretraining strategies described in [41]. The reported experimental results were averaged over the 5 trials, with evaluation based on the minimum phoneme-level MSE.

**Model Configuration.** The phoneme-level, word-level, and the utterance-level encoder (viz. $\mathrm{PhnEnc}(\cdot)$, $\mathrm{WordEnc}(\cdot)$, $\mathrm{UttEnc}(\cdot)$) consisted of 3, 2, and 1 convolution-augmented Branchformer blocks, respectively [25]. Within each encoder block, the self-attention branch was implemented with a single-head attention layer, followed by two feed-forward layers. Both the self-attention and feed-forward layers had a hidden dimension of 24. Meanwhile, the convolutional branch consisted of a depth-wise convolutional layer with a $1\times3$ kernel and a point-wise convolutional layer with a $1 \times 1$ kernel, both of which had 24 channels. To aggregate word-level and utterance-level features, the attention pooling modules were composed of a depth-wise convolutional layer and a single-head self-attention layer, where the convolutional layer had 24 channels with a kernel size of $1 \times 3$ and the attention layer had a hidden dimension of 24. Furthermore, we set the hidden dimension of the projection layers (viz. $\mathrm{Linear}_p(\cdot)$, $\mathrm{Linear}_w(\cdot)$, and $\mathrm{Linear}_u(\cdot)$) to 24. During the training phase, the tunable parameters of $\mathcal{L}_p$, $\mathcal{L}_w$, and $\mathcal{L}_u$ in Eq. (14) were set to 3, 1, and 1, respectively, while the temperature factor $\tau$ in Eqs (20) and (21) was set to 1.

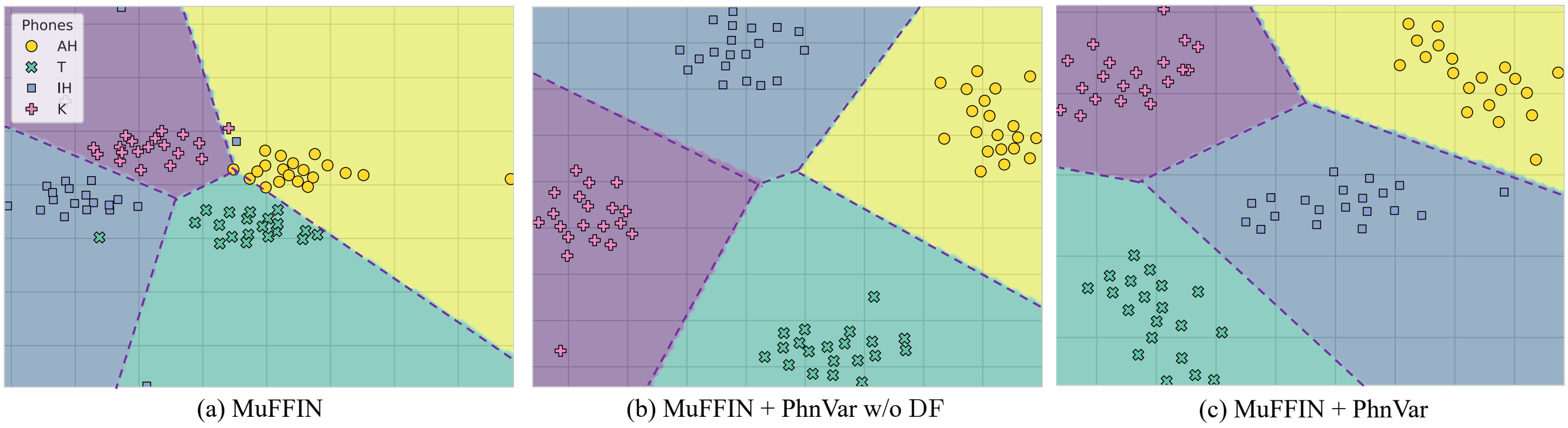

**Fig. 9.** Visualization of phoneme-level logits and decision boundaries from the diagnosis predictor of MuFFIN models trained with the proposed phoneme-specific variation (PhnVar) scheme. We show the phoneme logits of diagnosis predictor for (a) the vanilla MuFFIN model, (b) MuFFIN + PhnVar w/o DF (i.e., the variant of PhnVar without accounting for the pronunciation difficulty factor), and (c) MuFFIN + PhnVar.

**Mispronunciation Detection and Diagnosis via MuFFIN.** To detect mispronunciation segments, MuFFIN follows a pronunciation scoring-based paradigm, where the outputs of the phoneme-level error detector serve as indicators of mispronounced segments. The phoneme segments are identified as mispronounced if the corresponding indicators exceed a predefined threshold. Subsequently, the detected mispronunciation segments are fed into the phoneme-level predictor to generate diagnostic results. To ensure consistency between the detector and predictor, we mask the canonical phonemes (i.e., the phonetic transcription of the text prompt) during the softmax computation of the predictor.

### *C. Compared Methods*

We compare the proposed model (MuFFIN) with three categories of pronunciation assessment models. 1) Single-aspect pronunciation assessment: Lin2021 is a hierarchical APA model which takes phoneme-level surface features as inputs and assesses accuracy scores at utterance-level [57]. Kim2022 relies on layer-wise contextual representations extracted from a pretrained acoustic model to measure oral skills in terms of fluency or prosody at the utterance-level [30]. 2) Multi-aspect and multi-granular pronunciation assessment: For the assessment models with parallel neural structures, GOPT and LSTM are prominent models, both of which consume a sequence of GOP features and generate a set of proficiency scores at the phoneme-, word-, and utterance-level simultaneously [45]. 3M augments the input features of GOPT with SSL-based features to capture supra-segmental pronunciation cues, while also integrating phonological features to enhance phoneme-level textual information [46]. As for hierarchical models, HierGAT devises a language-hierarchy aware model with a series of graph attention neural networks and further strengthening the relatedness among the aspects with aspect attention mechanisms [58]. 3MH, a previous state-of-the-art method for APA, employs 3M as the backbone model and introduces sup-phoneme modeling to capture finer articulation traits within the language hierarchy between the phoneme and word levels [48]. Gradformer (GFR) decouples the language hierarchy into two sub-levels, i.e., lower (phoneme and word), and higher (utterance). A Conformer encoder models aspects at the lower linguistic level, while a Transformer decoder processes a sequence of learnable aspect vectors and interacts with the encoder outputs to assess aspect at the utterance-level [42]. 3) Multi-faceted pronunciation assessment: Ryu2023 introduces a unified model architecture that jointly optimizes both phoneme recognition and pronunciation assessments by independently stacking a CTC-based phoneme recognizer and a set of regressors on top of a pretrained acoustic model [59]. JAM advances 3M by integrating a phoneme classifier to predict diagnostic phonemes based on input canonical phonemes and further boosts the MDD performance by exploiting electromagnetic articulography (EMA) features to capture the articulatory movements of L2 learners [60].

## V. Experimental Result

### *A. Qualitative Analysis*

At the outset of experiments, we first analyze the phoneme statistics of Speechocean762 to reveal the intricate relationships between data quantity and pronunciation difficulty for each phoneme. Following this, we examine the effectiveness of the proposed phoneme-level regularizers through a series of qualitative visualizations.

**Phoneme Statistics of Speechocean762.** In Fig 6, the occurrence counts of phoneme segments (blue bars) paired with their respective mispronunciation rates (orange points) are

TABLE III
PERFORMANCE EVALUATION OF MUFFIN AND COMPARATIVE METHODS ON THE APA TASK OF SPEECHOCEAN762

| Model | Phoneme-level Acc. | | Word-level Aspects | | | Utterance-level Aspects | | | | |
|---|---|---|---|---|---|---|---|---|---|---|
| | MSE↓ | PCC↑ | Acc.↑ | Stress↑ | Total↑ | Acc.↑ | Comp.↑ | Fluency↑ | Prosody↑ | Total↑ |
| Lin2021 [57] | - | - | - | - | - | - | - | - | - | 0.720 |
| Kim2022 [30] | - | - | - | - | - | - | - | 0.780 | 0.770 | - |
| LSTM [45] | 0.089 (0.000) | 0.591 (0.003) | 0.514 (0.003) | 0.294 (0.012) | 0.531 (0.004) | 0.720 (0.002) | 0.076 (0.086) | 0.745 (0.002) | 0.747 (0.005) | 0.741 (0.002) |
| GOPT [45] | 0.085 (0.001) | 0.612 (0.003) | 0.533 (0.004) | 0.291 (0.030) | 0.549 (0.002) | 0.714 (0.004) | 0.155 (0.039) | 0.753 (0.008) | 0.760 (0.006) | 0.742 (0.005) |
| 3M [45] | 0.078 (0.001) | 0.656 (0.005) | 0.598 (0.005) | 0.289 (0.033) | 0.617 (0.005) | 0.760 (0.004) | 0.325 (0.141) | 0.828 (0.006) | 0.827 (0.008) | 0.796 (0.004) |
| GFR [42] | 0.079 (0.001) | 0.646 (0.004) | 0.598 (0.006) | 0.334 (0.013) | 0.614 (0.006) | 0.732 (0.005) | 0.318 (0.139) | 0.769 (0.006) | 0.767 (0.004) | 0.756 (0.003) |
| HierGAT [58] | 0.073 (0.001) | 0.683 (0.004) | 0.648 (0.003) | 0.327 (0.011) | 0.663 (0.002) | 0.798 (0.002) | 0.531 (0.047) | 0.840 (0.002) | 0.833 (0.002) | 0.821 (0.002) |
| 3MH [48] | 0.071 (0.001) | 0.693 (0.004) | 0.682 (0.005) | **0.361** (0.098) | 0.694 (0.007) | 0.782 (0.003) | 0.374 (0.115) | **0.843** (0.003) | **0.836** (0.004) | 0.811 (0.004) |
| Ryu2023 [59] | - | - | - | - | - | 0.719 | - | 0.775 | 0.773 | 0.743 |
| JAM [60] | 0.076 (0.002) | 0.664 (0.001) | 0.622 (0.012) | 0.241 (0.034) | 0.638 (0.005) | 0.773 (0.007) | 0.205 (0.080) | 0.831 (0.004) | 0.829 (0.004) | 0.805 (0.004) |
| MuFFIN | **0.063** (0.002) | **0.742** (0.006) | **0.705** (0.004) | 0.315 (0.033) | **0.714** (0.004) | **0.807** (0.003) | **0.768** (0.049) | 0.841 (0.004) | 0.832 (0.004) | **0.830** (0.002) |

* The reported results include the mean PCC scores and standard deviations, calculated over 5 independent experimental trials. Acc. and Comp. refer to the pronunciation aspects of accuracy and completeness, respectively. The proposed MuFFIN achieves higher PCC scores compared to 3MH across all metrics except utterance-fluency, with approximate randomization test ($p < 0.001$).

reported for the speechocean762 dataset, where the phonemes are sorted by mispronunciation rate and then categorized into three disjoint subsets: high (mispronunciation rate above 5.1%), medium (mispronunciation rate between 5.1% and 3.4%), and low (mispronunciation rate below 3.4%) regions.

In Fig 6, it is evident that the occurrence counts of phonemes and their corresponding mispronunciation rates exhibit distinct distributional patterns. For example, the high-occurrence phonemes (e.g., /AH/, /T/, and /N/) are found within the medium mispronunciation rate region. In contrast, some low-occurrence phonemes (e.g., /ZH/, /TH/, and /NG/) are often associated with high mispronunciation rates. Building on this, to mitigate the data imbalance issue facing the MDD task, the proposed phoneme-specific variance incorporates two novel regulation terms: a quantity factor and a pronunciation difficulty factor. The former balances the feature distributions of phonemes, while the latter adjusts feature scatteredness according to the mispronunciation rate.

**Qualitative Visualizations of Phoneme Representations for the Phonemic Characteristic and the Ordinal Terms.** In the second set of experiments, we graphically examining the impacts of the phonemic characteristic term and the ordinal term (i.e., $\mathcal{L}_{pc}$ and $\mathcal{L}_{o}$) based on the proposed APA model. As depicted Fig. 7, we extract the phoneme representations $\mathrm{H}^{p}$ from the test set and visualize each data point pertaining to the phoneme category (denoted by shape) and the corresponding pronunciation accuracy score (represented by color).

TABLE IV
PERFORMANCE EVALUATION OF MUFFIN WITH PHONEME-LEVEL REGULARIZERS UNDER THE PHONEME-SPECIFIC VARIATION TRAINING SCHEME

| MuFFIN | | | Phoneme Score | Word-level Score | | |
|---|---|---|---|---|---|---|
| PhnVar | $\mathcal{L}_{con}$ | $\mathcal{L}_{pco}$ | Acc. | Acc. | Stress | Total |
| - | - | - | 0.742 | 0.705 | 0.315 | 0.714 |
| V | - | - | 0.746 | 0.704 | 0.310 | 0.714 |
| V | V | - | **0.749** | 0.707 | 0.314 | **0.718** |
| V | - | V | 0.745 | 0.703 | 0.296 | 0.713 |
| V | V | V | 0.747 | **0.708** | **0.341** | **0.718** |

*Acc. refer to the pronunciation aspects of accuracy.

From Fig. 7(a), it is observed that despite MuFFIN jointly optimizing both the phoneme recognition and the assessment tasks, the resulting phoneme representations, however, are inevitably grouped by phoneme-level accuracy scores, inadequate to explicitly capturing the subtle distinctions between phonemes in the feature space. When training MuFFIN with $\mathcal{L}_{pc}$, as shown in Fig. 7(b), the phoneme-discriminative features are obtained, where the representations disperse according to their respective phoneme categories. However, simply separating the feature representations would omit the ordinal relationships, which might impede pronunciation assessment tasks. In response to this, the synergy of $\mathcal{L}_{pc}$ and $\mathcal{L}_o$ serves as a remedy, which enables the phoneme representations to reflect both categorical distinctions and ordinal relationships derived from their accuracy scores, as shown in Fig. 7(c). Specifically, integrating $\mathcal{L}_o$ leads to a stronger correlation between pairwise distances and phoneme-level accuracy within each phoneme category, resulting in an outward dispersion in the feature space as accuracy decreases. Grounded on these observations, incorporating $\mathcal{L}_{pc}$ and $\mathcal{L}_o$ during the training process of MuFFIN substantially improves the discriminability of phoneme representations and simultaneously reflects the ordinal relationships of the predicted accuracy scores in the feature space.

**Qualitative Visualizations of Phoneme Representations for the Proposed Contrastive Term.** Subsequently, to qualitatively assess whether the contrastive term $\mathcal{L}_{con}$ aligns the speech-derived representations (colored in blue) with their corresponding textual embeddings (colored in orange) for phoneme segments, we visualized the representations $\mathrm{H}^p$ and $\mathrm{E}^p$ from MuFFIN on the test set in Fig 8. By comparing among Fig. 8(a) and Fig. 8(b), we observe that the proposed $\mathcal{L}_{con}$ effectively projects these two types of phoneme representations into a shared feature space, resulting in a more coherent distribution. Going one step further, a zoomed-in view is presented in Fig. 8(c), which highlights that the contrastive term not only aligns the heterogeneous phoneme representations with the corresponding textual embeddings, but also preserves the phoneme-specific characteristics across phoneme categories.

**Qualitative Visualizations of Phoneme Logits for the Proposed Phoneme-specific Variation.** Finally, to qualitatively evaluate the effectiveness of the proposed PhnVar training scheme, we visualize the phoneme logits and decision boundaries of the diagnosis predictor. As shown in Figure 9, we compare the MuFFIN models trained with PhnVar and the variant (i.e., PhnVar without accounting for the pronunciation difficulty factor term). Furthermore, the visualized phonemes (/AH/, /T/, /IH/, and /K/) are uniformly sampled from the test set, with occurrence counts of 4.4K, 4K, 3K, and 1.3K, and mispronunciation rates of 4.80%, 3.55%, 5.46%, and 2.39%, respectively.

The observations from Fig. 9 are highlighted as follows. First, the logits of phonemes with higher occurrence counts tend to occupy a larger portion of the feature space, while those with lower occurrence counts are compressed into a narrower region. This is evidenced by the increasing size of feature regions for phonemes /K/, /IH/, /T/, to /AH/, which is consistent with their respective occurrence frequencies. Subsequently, in Fig. 9(b), it is observed that training MuFFIN with the variant of PhnVar (viz. PhnVar w/o DF) results in more uniformly distributed feature regions, independent of phoneme occurrence counts. However, adjusting the feature space solely factoring in data quantity factor fails to capture the distribution of mispronunciations. In light of this, our PhnVar additionally takes the pronunciation difficulty factor into account. The phoneme logits of MuFFIN trained with PhnVar are visualized in Fig. 9(c), where the feature regions are partitioned by the phoneme mispronunciation rates, with region sizes decreasing in the order of /IH/, /AH/, /T/, and /K/.

### B. *Performance of Automatic Pronunciation Assessment*

In this subsection, we turn to evaluating the performance of MuFFIN on pronunciation assessments and compare it with several state-of-the-art models to validate its effectiveness. The corresponding results are presented in Table III. We begin by discussing the experimental results at the phoneme- and word-level assessments and then proceed to the utterance-level assessments.

**Assessment Performance at Phoneme and Word Levels.** We first evaluate the assessment performance at the phoneme and word levels in Table III, from which the following observations can be made. First, the proposed MuFFIN outperforms other APA models by a remarkable margin in most pronunciation assessment tasks, except for the word-level stress. Specifically, MuFFIN stands out in the phoneme-level accuracy, demonstrating PCC score improvements of 4.9% and 5.9% over the prior-art models, 3MH and HierGAT, respectively. We attribute these performance gains to the proposed multi-faceted phoneme-level pronunciation feedback module, which jointly optimizes the APA and MDD tasks, thereby encouraging the phoneme encoder to learn distinct phoneme identities when evaluating the pronunciation scores. With respect to the word-level assessments, MuFFIN generally performs well across most pronunciation aspects. However, in word-level stress, our model demonstrates comparable performance against GFR and HierGAT, while trailing behind 3MH. A possible reason for the inferior performance is that 3MH leverages sub-phoneme modeling to create a pseudo (augmented)-linguistic hierarchy between phoneme and word levels, facilitating better rendering of supra-segmental information for word-level assessments.

Second, we turn to evaluate the performance of strong baselines with parallel neural architectures (the second group in Table III). Compared to LSTM and GOPT, 3M stands out as a promising method, with its superiority stems from effectively exploiting SSL-based features to mitigate the data scarcity issue of L2 learners' speech and simultaneously encapsulate long-range articulatory traits. By augmenting the inputs of 3M with electromagnetic articulography features, JAM slight boosts the performance in most pronunciation assessment tasks. Finally, among the APA models with advanced neural architectures (the third group in Table III), 3MH achieves the best performance, benefiting from the synergy of hierarchical modeling approaches and depth-wise convolution layers. However, compared to MuFFIN, 3MH is limited in functionality, as it only qualifies pronunciation proficiency with various aspect scores, lacking phoneme-level diagnostic feedback.

**Assessment Performance at Utterance-level.** For the performance of utterance-level pronunciation assessment in Table III, MuFFIN achieves the highest performance across most aspects. Compared to 3MH, MuFFIN enhances the PCC scores by 2.5% in utterance-level accuracy, 2.9% in utterance-level total, and achieves comparable performance in utterance-level fluency and prosody. MuFFIN also achieves substantial improvements in the utterance-level completeness assessment, a metric reflecting the proportion of correctly pronounced words in an utterance. This gain is attributed to the joint training

TABLE V
PERFORMANCE EVALUATION OF MUFFIN AND COMPARATIVE METHODS ON MDD TASK

| Model | Mispronunciation Detection | | | Mispronunciation Diagnosis | | | PER (%)↓ |
|---|---|---|---|---|---|---|---|
| | RE (%)↑ | PR (%)↑ | F1 (%)↑ | FAR (%)↓ | FRR (%)↓ | DER (%)↓ | |
| Ryu2023 [59] | **91.60** | 26.90 | 41.50 | - | - | - | 9.93 |
| JAM [60] | 34.76 | 61.10 | 45.01 | 64.32 | **0.58** | **45.23** | 2.81 |
| MuFFIN | 64.33 | 66.89 | 65.99 | 35.67 | 0.97 | 60.97 | 2.36 |
| w/ PhnVar | 68.37 | **67.60** | **67.98** | **31.63** | 1.01 | 58.82 | **2.33** |

* For the mispronunciation detection subtask, our best-performing model (MuFFIN+PhnVar) outperforms the base model (MuFFIN) with significantly better performance ($p < 0.001$).

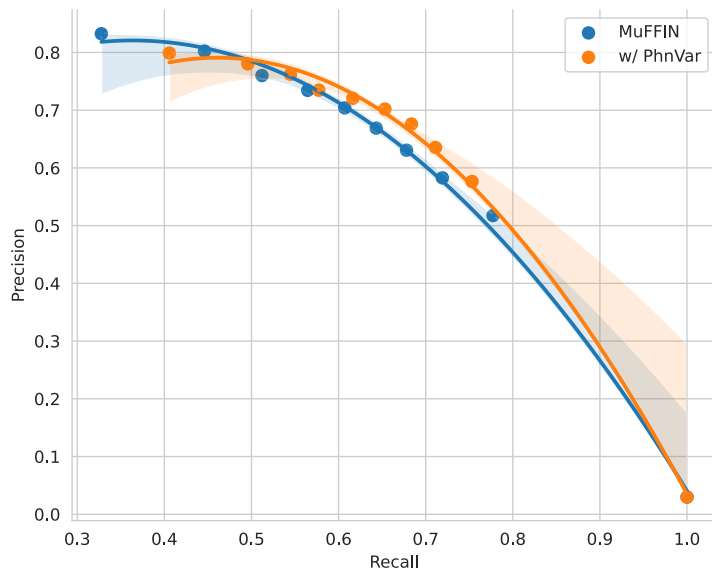

**Fig. 10.** Precision-recall curves for MuFFIN trained with PhnVar.

of the MDD task within the APA model, which consequently enables MuFFIN to pinpoint mispronounced segments and identify corresponding phonemes in learners' speech. By leveraging phoneme-discriminative representations, MuFFIN effectively propagates fine-grained information from phoneme- to utterance-level assessments through a tailored, hierarchy-aware neural architecture.

Next, compared to other strong baseline models, Lin2021 trails behind several APA models trained on multiple pronunciation aspects (the second group in Table III), revealing that the single-aspect assessment models fail to leverage the dependency relationships between aspects through multi-task learning, thereby leading to inferior performance. In subsequent work, Kim2022 ventures into replacing conventional ASR-driven features with SSL-based features, resulting in substantial improvements and achieving comparable performance to 3M. In comparison with multi-faceted pronunciation assessment models (viz. JAM and Ryu2023), JAM demonstrates superior performance in most aspects of the utterance-level assessment. This improvement may stem from the novel use of fine-grained phoneme-level features, including GOP features, prosody statistics, and EMA features.

**Effectiveness of Phoneme-specific Variation and ConPCO.** Lastly, we examine the effectiveness of the proposed training scheme, phoneme-specific variation (PhnVar), and the contrastive phonemic ordinal regularizer (ConPCO) for the pronunciation assessment. In Table IV, we concentrate on the assessment performance at both the phoneme and word levels, as we empirically found that the proposed regularizers do not cause detrimental effects on the utterance-level pronunciation assessments, with performance either slightly improved or at least on par with the vanilla MuFFIN model [24][25]. In addition, ConPCO is decomposed into the contrastive term ($\mathcal{L}_{con}$) and the phonemic ordinal regularizer ($\mathcal{L}_{pco} = \mathcal{L}_{pc} + \mathcal{L}_{o}$), both of which are then combined with PhnVar for training MuFFIN.

From Table IV, we can observe that the proposed training scheme, PhnVar, the proposed PhnVar training scheme yields a 0.7% improvement in phoneme-level accuracy over the base model. Subsequently, the incorporation of the phoneme-level regularizers (viz., $\mathcal{L}_{con}$ and $\mathcal{L}_{pco}$) under the PhnVar training regime benefits pronunciation assessments, as evidenced by the sustained or improved results at the phoneme- and word-level assessment tasks. Furthermore, the contrastive term primarily boosts the performance in the aspects of phoneme-level accuracy and word-level total score. In contrast, the phonemic ordinal regularizer tends to either slightly enhance performance or retain that of the vanilla MuFFIN model. In addition, training MuFFIN with ConPCO attains the best performance in the word-level assessment tasks (as shown in the last row of Table IV).

### C. *Performance of Mispronunciation Detection and Diagnosis*

In this subsection, we evaluate the performance of MDD for the multi-faceted pronunciation assessment models. As joint optimization of APA and MDD remains relatively underexplored in CAPT, only a limited number of relevant studies are available for comparison in the following experiments. To the best of our knowledge, Ryu2023 is the first attempt to develop a multi-faceted pronunciation feedback model, while JAM represents the recent follow-up work. The main results of the MDD task are summarized in Table V. Next, Tables VI delve deeper into the imbalance issues inherent in the MDD, demonstrating the effectiveness of the proposed phoneme- variation training (PhnVar) scheme.

**Performance Evaluation of MDD.** To detect mispronunciations with the MuFFIN, we leverage the outputs of the phoneme-level error detector to identify mispronunciation segments. Specifically, we first collect the detector outputs for all phoneme segments in the training set. A global threshold is then selected through grid search, with a stride of 0.1 over a range $[0.0, 1.0]$. At the outset of MDD experiments, we present the precision-recall curves of MuFFIN and the MuFFIN trained with PhnVar in Fig. 10. For our models (i.e., MuFFIN and MuFFIN+PhnVar) this global threshold is set to 0.4. In contrast to prior dictation-based methods (e.g., Ryu2023 and JAM), which detect mispronunciations by identifying discrepancies between recognized and canonical phonemes, MuFFIN adopts a scoring-based approach that detects mispronunciations via threshold tuning, facilitating broader adaptability to diverse L2 learners.

As shown in Table V, MuFFIN outperforms other methods in the mispronunciation detection subtask, achieving outstanding performance in terms of F1-score and precision. Moreover, training MuFFIN with the phoneme-specific variation (PhnVar) leads to notable improvements in all evaluation metrics compared to the base model. This gain is further illustrated in Fig. 10, where the orange line (MuFFIN+PhnVar) exceeds the blue line (MuFFIN) in area under the precision–recall curve. Subsequently, in comparison with other baseline methods, Ryu2023, on the basis of a CTC-

TABLE VI
PROBING THE IMBALANCE ISSUES IN MDD WITH MUFFIN

| Group | Metrics | Average PER (%)↓ | | | Average Recall (%)↑ | | | Average Precision (%)↑ | | | Average F1-score (%)↑ | | |
|---|---|---|---|---|---|---|---|---|---|---|---|---|---|
| Occurrence Count | Type | Many | Med. | Few | Many | Med. | Few | Many | Med. | Few | Many | Med. | Few |
| | MuFFIN | 1.19 (0.31) | 2.27 (0.14) | 10.93 (25.79) | 70.19 (7.76) | 75.53 (15.61) | 68.56 (22.75) | 50.84 (12.31) | 53.42 (19.05) | 69.50 (24.79) | 57.53 (6.95) | 60.14 (14.59) | 67.77 (21.43) |
| | PhnVar | 1.41 (0.34) | 2.41 (1.51) | 9.38 (20.19) | 61.90 (6.22) | 64.04 (17.41) | 68.97 (24.66) | 66.82 (9.81) | 66.27 (16.03) | 72.70 (24.14) | 63.86 (5.80) | 62.71 (10.64) | 69.05 (9.38) |
| | w/o DF | 1.45 (0.34) | 2.42 (1.63) | 11.22 (25.83) | **64.71 (8.43)** | **68.62 (23.62)** | **67.99 (24.77)** | 61.27 (12.99) | 58.87 (14.96) | 61.83 (22.68) | **63.02 (7.94)** | 59.23 (14.34) | 62.38 (19.55) |
| | w/o QF | **1.28 (0.39)** | **2.09 (1.36)** | **9.89 (21.73)** | 57.55 (6.39) | 62.72 (22.32) | 60.02 (23.27) | **69.22 (9.90)** | **64.04 (19.84)** | **76.22 (24.85)** | 62.58 (6.80) | **61.23 (18.60)** | **65.22 (20.65)** |
| Mispron. Rate | Type | High | Med. | Low | High | Med. | Low | High | Med. | Low | High | Med. | Low |
| | MuFFIN | 10.62 (25.82) | 1.69 (1.01) | 2.08 (2.41) | 77.43 (12.19) | 69.12 (8.91) | 67.73 (23.77) | 61.70 (16.77) | 55.83 (20.91) | 56.23 (24.37) | 66.95 (10.94) | 59.81 (14.18) | 58.68 (20.39) |
| | PhnVar | 9.30 (20.13) | 1.82 (0.91) | 2.15 (2.36) | 70.51 (14.30) | 62.19 (10.57) | 62.22 (24.78) | 70.88 (12.35) | 67.97 (16.37) | 66.95 (23.08) | 69.10 (6.93) | 64.22 (10.94) | 62.30 (20.47) |
| | w/o DF | 10.86 (25.75) | 1.80 (0.92) | 2.42 (3.68) | **71.33 (12.52)** | **73.52 (16.62)** | **59.16 (26.29)** | 66.10 (14.05) | 54.20 (14.12) | 61.67 (21.04) | 66.77 (8.10) | 60.79 (10.93) | 57.08 (20.64) |
| | w/o QF | **9.27 (21.65)** | **1.70 (1.06)** | **2.29 (3.72)** | 66.23 (14.28) | 55.39 (16.96) | 58.67 (23.25) | **71.94 (12.39)** | **70.43 (21.93)** | **67.11 (23.18)** | **67.29 (8.42)** | **61.04 (17.27)** | **60.69 (20.74)** |

*Performance gains in the comparison of PhnVar variants (w/o DF and w/o QF) are highlighted in bold, while improvements over the vanilla MuFFIN achieved by training with PhnVar are marked with underlines. 'Mispron. Rate' denotes the mispronunciation rate.

based phoneme recognizer, achieves the highest recall value but has the downside of low precision for the mispronunciation detection task. This limitation is consistent with the shortcomings reported for dictation-based MDD models [37][38], where model performance is intrinsically constrained by the phoneme recognition rate. Instead of a direct free-phoneme recognition process, JAM builds upon 3M and detects mispronunciations in learners' speech by attaching a phoneme classifier to the phoneme-level encoder. The corresponding result demonstrates promising performance in terms of precision metric, though it struggles with the low recall rate. Compared to JAM, our MuFFIN achieves superior performance across all metrics in the mispronunciation detection subtask. These findings collaboratively highlight the effectiveness of the proposed scoring-based approach for multi-faceted pronunciation feedback.

For mispronunciation diagnosis subtask, our methods achieve promising performance in terms of FAR and PER. However, a trade-off appears to exist between recall and the metrics of FRR and DER. Specifically, compared to JAM, MuFFIN achieves higher recall rate and lower PER but exhibits inferior performance in both FRR and DER. This result implies that our model detects a greater number of mispronounced segments but comes at the cost of diagnostic accuracy. We leave this issue as a direction for future research.

**Systematic Examination of the Data Imbalance Problem in MDD.** In the following section, we explore the data imbalance problem in MDD, which stems from two intertwined factors within the phoneme segments, i.e., data quantity and pronunciation difficulty. To disentangle these two factors, we divide the phoneme segments into two groups and report the corresponding phoneme error rate (PER), recall, precision, and F1-score for each subset, with their mean and standard deviation. As shown in Table VI, in the first group, phoneme segments are categorized into many (occurrence count above 1.3K), medium (occurrence count between 1.3K and 0.6K), and few (occurrence count below 0.6K) shot regions based on their occurrence counts. Conversely, in second group, phoneme segments are categorized into high (mispronunciation rate above 5.1%), medium (mispronunciation rate between 5.1% and 3.4%), and low (mispronunciation rate below 3.4%) pronunciation error regions according to their mispronunciation rates (*cf.* Fig. 6). Furthermore, this set of experiments is conducted to assess the upper-bound performance of MuFFIN, with the aims of analyzing the imbalance issue of MDD and examining the effectiveness of the proposed phoneme-specific variance (PhnVar). To this end, a held-out set of 500 utterances is set aside from the training data, with the remaining 2,000 utterances used for model training. This held-out set is designed to cover both correct and incorrect pronunciations of each phoneme and is then used to determine phoneme-specific thresholds by maximizing the area under the precision–recall curve. Based on these phoneme-specific thresholds, we then evaluate the MDD performance on the Speechocean762 test set.

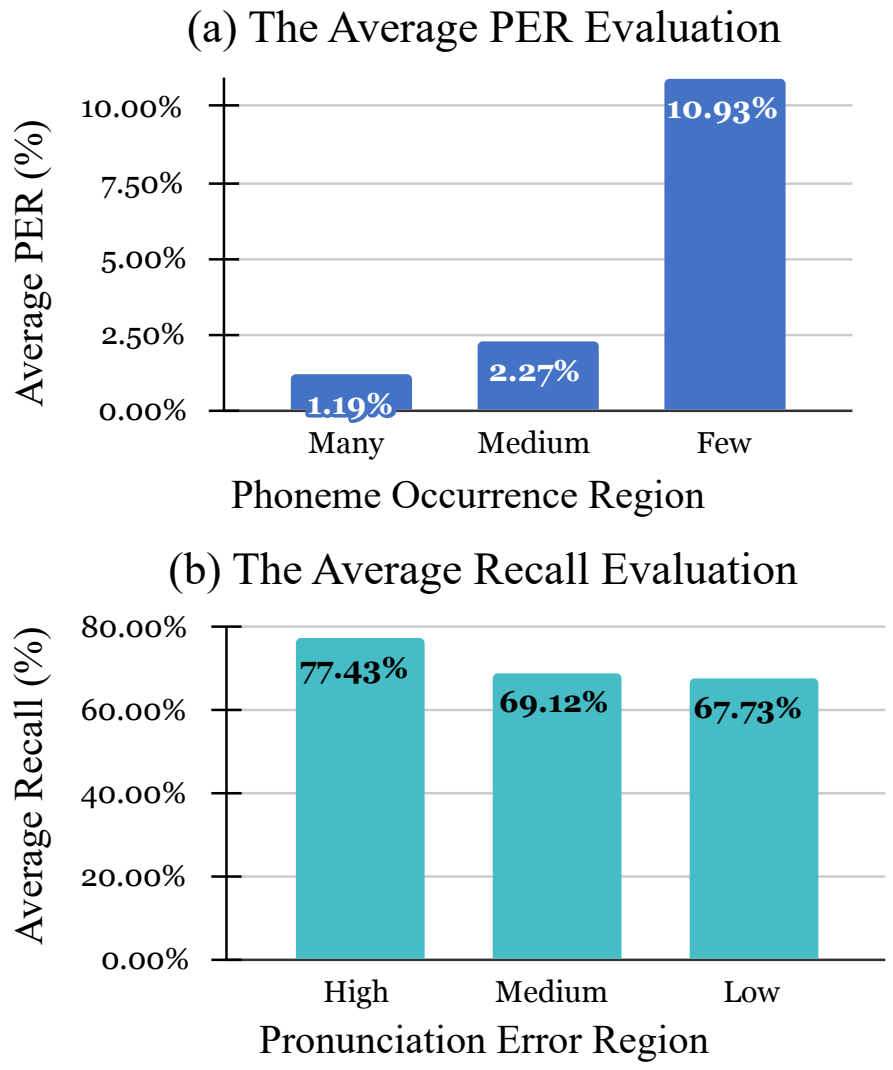

**Fig. 11.** The performance evaluations highlight data imbalance issues in MDD. Our bar charts demonstrate that (a) the data quantity factor primarily affects average PER in phoneme subsets grouped by occurrence count, and (b) the pronunciation difficulty factor significantly influences average recall in phoneme subsets grouped by mispronunciation rate.

To highlight the data imbalance issues, Fig. 11 first presents two sets of bar charts based on the MuFFIN model, where Fig. 11(a) displays the average PER across phoneme subsets grouped by occurrence count and Fig. 11(b) shows the average recall for those grouped by mispronunciation rate. In Fig. 11(a), we can observe that the average PER of MuFFIN increases significantly from many-shot to few-shot region. This observation is consistent with findings from prior studies on data imbalance learning [22][23], which manifests that occurrence frequency figures prominently in phoneme recognition accuracy. A naïve training process for a phoneme classifier based on empirical risk minimization inevitably biases the model toward majority phoneme categories, resulting in inferior performance on minority categories. Apart from the data quantity issue, Fig. 11(b) reveals that the pronunciation difficulty factor causes a steady decline in average recall as phoneme subsets shift from high to low mispronunciation rates. This suggests that the infrequently mispronounced phoneme segments pose greater challenges for pronunciation error detection. Drawing on these observations, the proposed PhnVar training scheme integrates two mathematical terms designed to simultaneously account for data quantity and pronunciation difficulty.

We next investigate the efficacy of each component in PhnVar for handling the imbalance issues of MDD. The ablation studies in Table VI are conducted by excluding either the data quantity factor (w/o QF) or the pronunciation difficulty factor (w/o DF) from the proposed PhnVar. As shown in the table, we observe a consistent performance trend across these two groups, where the MuFFIN model gains in recall by considering the data quantity factor (w/o DF), while precision benefits from incorporating the pronunciation difficulty factor (w/o QF). Subsequently, by factoring in both factors, PhnVar achieves notable performance gains in F1-score. These results shed light on the fact that simply balancing the logits of phoneme predictions encourages the MDD model to detect a greater number of pronunciation errors. This, however, boosts the recall rate while resulting in a decrease in precision evaluation. A plausible explanation is that the mispronunciation rates are not uniformly distributed across all phonemes. As a remedy, the proposed PhnVar takes both the factors of data quantity and the pronunciation difficulty into account, striking a balance between recall and precision to achieve an optimal F1-score. Furthermore, compared to the vanilla MuFFIN model, training with PhnVar boosts the performance significantly in precision and F1-score, with gains particularly evident for phoneme subsets with medium and few occurrences, as well as those with medium and low mispronunciation rates.

TABLE VII
ABLATION STUDY ON THE VARIED GRANULARITIES IN TRAINING OBJECTIVES OF APA FOR MUFFIN

| Training Objective | Number of Params | Pronunciation Aspects (PCC Score) | | |
|---|---|---|---|---|
| | | Phone Accuracy | Word Accuracy | Utt. Accuracy |
| Utt. Only | 541K | - | - | 0.782 |
| Word Only | 248K | - | 0.674 | - |
| Phone Only | 126K | 0.715 | - | - |
| +Word | 249K | 0.724 | 0.688 | - |
| +Utt. | 608K | **0.726** | **0.687** | **0.807** |

*In this table, MuFFIN refers to training without MDD task. 'Utt.' denotes the utterance.

TABLE VIII
ABLATION STUDIES ON TRAINING GRANULARITIES FOR MUFFIN WITH APA AND MDD OBJECTIVES

| Training Objective | APA Task | | | MDD Task | | |
|---|---|---|---|---|---|---|
| | Phone Acc. | Word Acc. | Utt. Acc. | F1 (%) | RE (%) | PR (%) |
| MDD Only | - | - | - | 62.71 | 65.67 | 60.33 |
| +Utt. | - | - | 0.787 | 63.34 | 63.49 | 63.45 |
| +Word | - | 0.681 | - | 64.46 | 66.27 | 62.86 |
| +Phone | 0.717 | - | - | **66.26** | **69.06** | 63.77 |
| +Word | 0.741 | 0.696 | - | 66.04 | 67.08 | 65.36 |
| +Utt. | **0.742** | **0.705** | **0.807** | 65.99 | 64.33 | **66.89** |

'Utt.' denotes the utterance and 'Acc.' denotes the accuracy aspect.

### *D. Ablation Studies for objectives of APA and MDD*

In this subsection, a series of ablation studies are carried out to analyze the effectiveness of various training objectives for MuFFIN on both pronunciation accuracy and mispronunciation detection performance.

**Effectiveness of Multi-granularity Pronunciation Assessments in MuFFIN.** In this set of experiments, we begin by training MuFFIN without the MDD task, then progressively incorporate pronunciation assessment tasks at different linguistic levels. Table VII reports on the PCC scores on assessment of pronunciation accuracy for MuFFIN with various training objectives, comprising the assessment tasks at phoneme, word, utterance levels, as well as cross-granularity combinations. From Table VII, we observe that MuFFIN trained in a multi-granularity manner achieves superior results in relation to any single-granularity assessment model compared in this paper, which indicates a strong correlation among assessment tasks within the linguistic hierarchy of an utterance. For instance, MuFFIN trained with multi-granularity objectives, i.e., Phone+Word and Phone+Word+Utt., outperforms their single-granularity counterparts, i.e., Word

Only and Utt. Only, with respective gains of 14% and 13%. Furthermore, a comparison of parameter sizes reveals that utterance-level assessment models (i.e., Utt. Only and Phone+Word+Utt.) have substantially larger parameter sizes than the other assessment models of different granularity combinations (e.g., Phone Only, Word Only, and Phone+Word). We attribute this to the residual connections between the mean pooling feature $\bar{\mathrm{E}}^{\mathrm{SSL}}$ and the utterance-level regressors.

**Effectiveness of Joint MDD and APA Training for MuFFIN.** We next ablate the individual contributions of the training objectives for both MDD and APA to the multi-faceted pronunciation assessment models. Table VIII details the performance of MuFFIN in jointly addressing the MDD and APA tasks, where the evaluation metrics are reported for both MDD and the PCC scores on pronunciation accuracy across various granularities. A closer look at Table VIII, we have the following observations. 1) Multi-faceted pronunciation models (e.g., MDD+Utt., MDD+Word, and MDD+Phone) that integrate APA tasks consistently outperform the model trained solely on MDD (viz. MDD Only) across all MDD evaluation metrics, demonstrating the synergistic effect of jointly modeling MDD and APA. 2) Among these multi-faceted pronunciation models, the model trained with phoneme-level assessment and MDD tasks (MDD+Phone) yields the optimum performance in term of the recall metric. 3) Regarding the performance of pronunciation assessment, observations from Tables VII and VIII suggest that the integration of MDD tasks maintains or slightly improves pronunciation accuracy. Finally, the primary improvement for the performance of pronunciation assessment stems from the incorporation of diverse assessment tasks at various linguistic levels.

## VI. Conclusion

In this paper, we have proposed a novel multi-faceted pronunciation feedback model dubbed MuFFIN which is designed to qualify learners' pronunciation from multiple perspectives, including pronunciation aspects across various linguistic levels, as well as mispronunciation detection and diagnosis at phoneme-level. A novel contrastive phonemic ordinal regularizer has been put forward to empower MuFFIN to generate more phoneme-discriminative features while accounting for the ordinal nature of phoneme-level accuracy scores. Furthermore, to tackle the intricate data imbalance problem of MDD, we present a simple yet effective training scheme that perturbs the outputs of a phoneme classifier with phoneme-specific variations. This approach effectively balances the distribution of predicted phonemes while incorporating considerations of pronunciation difficulty. The practical utility of our method has been verified through extensive experiments on speechocen762 benchmark dataset. The proposed contrastive phonemic ordinal regularizer has been thoroughly examined through a series of graphical visualizations. Moreover, this study is the first attempt to address the data imbalance in MDD from the perspectives of data quantity and pronunciation difficulty. The empirical results demonstrate that our model outperforms some state-of-the-art methods in both APA and MDD tasks.

**Limitations and Future Work.** The proposed method is constrained by its dependence on the "read-aloud" learning scenario and to some extent lacks explainability for the provided assessment results. Such scripted-speech assessments fail to reflect learners' speaking abilities in real-world communication. Furthermore, the experimental dataset solely contains the Mandarin learners, potentially hindering the generalization abilities and applicability to learners with other accents. In future work, we plan to examine the proposed method on spoken language assessment, where learners speak freely or respond to a given task or question [61]. In addition, the issues of explainable pronunciation feedback are also left as a future extension.